\documentclass[12pt]{article}
\usepackage{graphicx}
 
\begin{document}

\begin{center}

{\Large\bf Asteroseismic  model of paramagnetic neutron star}

\vspace{1.cm}

S.I. Bastrukov$^{1,2}$, I.V. Molodtsova$^{2}$, H.-K. Chang$^{1}$, J. Takata$^{3,4}$

\vspace{1.cm}

$^1$Institute of Astronomy,
  National Tsing Hua University, Hsinchu,  Taiwan\\
$^2$Joint Institute for Nuclear Research,  Dubna, Russia\\
$^3$ASIAA/National Tsing Hua University - TIARA, Hsinchu, Taiwan\\
$^4$Department of Physics, Hong Kong University, Hong Kong, China

\end{center}

\begin{abstract}
 We investigate an asteroseismic model of non-rotating paramagnetic neutron star with
 core-crust stratification of interior pervaded by homogeneous internal and dipolar external magnetic field,
 presuming that neutron degenerate Fermi-matter of the star core is in the state of Pauli's paramagnetic permanent magnetization caused by polarization of spin magnetic moments of neutrons along the axis of magnetic field of collapsed massive progenitor. The magnetic cohesion between metal-like crust and permanent-magnet-like core is considered as playing a main part in the dynamics of starquake. This process is thought of as impulsive release of magnetic field stresses, brought about by disruption of magnetic field lines on the core-crust interface resulting in the crust fracture, which is followed by tectonic shear displacements of crust against massive core. Focus is laid on the post-quake relaxation of the star by node-free torsional vibrations of highly
 conducting crustal solid-state plasma, composed of nuclei embedded in the degenerate Fermi-gas of relativistic electrons, about axis of magnetic field frozen in the immobile paramagnetic core. Two scenarios of these axisymmetric seismic vibrations are examined, in first of which these are considered as
 maintained by combined action of Lorentz magnetic and Hooke's elastic forces and in second one
 by solely Lorentz force. The damping is attributed to shear viscosity of crustal material. Based on the energy variational method of magneto-solid-mechanical theory of elastic continuous medium, the spectral formulae for the frequency and lifetime of this toroidal mode are obtained and discussed in the context of theoretical treatment of recently discovered quasi-periodic oscillations of the X-ray outburst flux from SGR 1806-20 and SGR 1900+14 as being produced by above seismic vibrations.

\end{abstract}



\section{Introduction}
 Studying seismic vibrations of neutron stars is among the most promising tools in search for dynamical laws governing continuum mechanics and macroscopic electrodynamics of extremely dense matter compressed by self-gravity to the nearly normal nuclear density at which microscopic composition of stellar material is dominated by neutron component. One of the most  conspicuous features distinguishing this subclass of final stage (FS) stars of evolutionary track from  their massive main-sequence (MS) progenitors is that they come into existence as sources of ultra powerful magnetic fields (Chanmugam 1992). It is hoped, therefore, that exploring vibrational behavior of spherical material masses capable of accommodating super strong
 magnetic fields we can gain some insight into the nature of the nuclear matter magnetism.

 In this connection it is appropriate to emphasize that contrary to liquid MS stars whose magnetic
 fields are generated in dynamo processes by flows of perfectly conducting flowing matter (energy supply of persistent convection in which comes from cental thermonuclear reactive zones), in the FS stars, white dwarfs and pulsars, there are no nuclear energy sources to  support persistent convection (e.g. Flowers \& Ruderman 1977, Spruit 2008).
 Such point of view invalidates applicability to FS stars of magneto-fluid-mechanical arguments and
 methods in the analysis of magnetic field effects such, for instance, as anti-dynamo theorems.
 Unlike the dynamo-generated fairly weak fields of liquid MS stars, the super strong and highly stable to spontaneous decay magnetic fields of isolated neutron stars (Bhattacharya \& van den Heuvel 1991, van den Heuvel 1994, Bhattacharya 2002) are regarded as fossil (e.g. Ferrario and Wickramasinghe 2004). These are thought of as
 frozen in the immobile, non-convective, degenerate stellar matter possessing
 properties of elastically deformable solid. It is believed, therefore, that basic features of asteroseimology of strongly magnetized end-products of stellar evolution (white dwarf stars, neutron stars and still hypothetical quark stars) can properly be understood working from equations of solid-mechanical theory of continuous media\footnote{It may be appropriate to mention here investigations in nuclear physics on theoretical treatment of giant resonances in terms of vibrations of ultra fine piece of continuous nuclear matter in which it has been found that behavior of atomic nucleus in these fundamental modes of nuclear response
 bears strong resembles to spheroidal and torsional vibrations of elastically deformable solid sphere rather than drop of flowing liquid. (e.g. Bastrukov et al 2008b and references therein). One of the remarkable analytic inferences of these investigations is that the shear modulus of nuclear material (which is treated, therefore, as an elastic Fermi-solid) entering equations of nuclear solid-globe model, is found to be precisely coinciding with the pressure of degenerate Fermi-system of nucleons. This finding of nuclear physics suggests that neutron star can be thought of
 as a spherical mass of an elastic Fermi-solid of the normal nuclear density in which the energy of
 gravitational pull is counterbalanced by elastic energy stored in the incessant Fermi-motion of neutron quasi-particles in the potential of self-gravity (e.g. Bastrukov et al 2007a).
 Bearing this in mind and widely spread opinion that a neutron star can be regarded as an astrophysical counterpart of an atomic nucleus (e.g. Lipunov 1992, Bisnovatyi-Kogan 2002), it has been shown in (Bastrukov et al 1999a, 1999b) that equations of nuclear elastodynamics, originally introduced in the nuclear physics of giant resonances as fundamental equations of continuum mechanics of nuclear matter, can be successfully utilized for the study of neutron star pulsations. And in recent work (Bastrukov et al 2007a), it was shown
  that linearized equations of nuclear elastodynamics and canonical equations of linear theory of elasticity
  lead to identical results regarding frequencies of
  fundamental modes of node-free spheroidal and torsional vibrations of spherical mass of viscoelastic solid which are of particular interest for our present discussion.} (McDermott et al 1988, Blaes et al 1989, Strohmayer 1991, Bastrukov 1996, Bastrukov et al 1999a, 1999b, Franco et al 2000, Bastrukov et al 2007 see, also, references therein).

 Many agree today that recent observations of quasi-periodic oscillations (QPOs) in the outburst X-ray flux of two soft gamma-ray repeaters, SGR 1806-20 and SGR 1900+14 (Israel et al 2005, Watts \& Strohmayer 2006)
 have offered a real opportunity of using the detected QPO frequencies for asteroseismological inferences about internal structure, mechanical and magnetic properties of material of these ultra magnetized neutron stars whose X-ray bursting activity exhibits many features in common with earth quakes (Chen et al 1996, Woods \& Thompson 2006, Mereghetti 2008). The detected QPOs frequencies (measured in Hz), which are of particular interest for the subject of present work, are
 \begin{eqnarray}
  \label{e1.1}
  && \mbox{\rm SGR}\, 1806-20: 18, 26, 29, 92, 150, 625, 1840;\\
  \label{e1.2}
  && \mbox{\rm SGR}\, 1900+14:  28, 54, 84, 155.
 \end{eqnarray}
 With all above in mind, in (Bastrukov et al 2007-2009) several models of post-quake vibrational relaxation of a solid star have been investigated.
 Particular attention has been given to the regime of node-free or nodeless shear axisymmetric vibrations which remain less investigated in  theoretical seismology as compared to the standing-wave regime of vibrations of both solid FS stars and such solid celestial objects as Earth-like planets (e.g. Lapwood \& Usami 1981, Lay \& Wallace 1995, Aki \& Richards 2003). In particular, in (Bastrukov et al 2007b, 2008a), a case of the elastic-force-driven nodeless shear oscillations, both torsional --  $_0t_\ell$ and spheroidal -- $_0s_\ell$, entrapped in the finite-depth crust $\Delta R$ of core-crust model of a solid star has been studied in some details with remarkable inference that dipole overtones of spheroidal and torsion nodeless elastic shear vibrations of crust against immobile core exhibit features generic to Goldstone soft modes. On the other hand one can casts doubt on arguments of model presuming that detected QPOs have been brought about by vibrations restored by solely solid-mechanical force of Hooke's elastic stresses because such interpretation rests on the poorly justifiable assumption of the dynamically passive role of ultra strong magnetic field, as defining only direction of the axis about which the stellar matter undergo differentially rotational fluctuations, i.e., that magnetic field frozen in the star remains unaltered in this process.
 Bearing this in mind and assuming that the admixture of charged particles (basically electrons and protons) imparts to neutron-dominated Fermi-matter of these star mechanical properties of perfectly conducting material,
 it seems more plausible to expect that post-quake relaxation should be strongly affected by Lorentz force of magnetic field stresses setting highly conducting material in rotational motion about axis of the dipole magnetic moment of the star.

  With this in mind, in (Bastrukov et al 2009a, 2009b), the above data have been analyzed in the model of a solid star
  undergoing global torsional nodeless vibrations about magnetic axis under the joint action of Lorentz force of magentic field stresses and  Hooke's force of elastic shear stresses. And it was found that such a model provides fairly reasonable account of general trends in data on QPO frequencies for SGR 1900+14 and data for SGR 1900+14 for frequencies from the range $30\leq \nu \leq 200$ Hz, but
  faces serious difficulties in interpreting the low-frequency vibrations with $\nu=18$ and $\nu=26$ Hz in the
  SGR 1806-20 data. Also, the model of global torsional vibrations of solid star with froze-in homogeneous magnetic field leaves some uncertainties regarding the nature of vibrations with $\nu=625$ and  $\nu=1840$ Hz. This last issue has been scrutinized in recent work (Bastrukov et al 2009b) from the standpoint of a solid star model with non-homogeneous poloidal magnetic field of well-known Ferraro's form and was found that these high-frequency QPOs can be properly interpreted as being produced by global Alfv\'en node-free torsional vibrations of very high overtones.

  In the meantime, it has been argued some time ago in the studies of pulsar glitches as being produced by star quakes (Bastrukov et al 1996, 1997, 1999b) that post-glitch relaxation should most likely be dominated by Alfv\'en oscillations of crustal electron-nuclear solid-state plasma (composed of immobile nuclei dispersed
  in degenerate Fermi-gas of relativistic electrons) about axis of magnetic field frozen in the immobile core
  (and, also, by less dense gaseous plasma expelled from the surface by starquake).
  From the view point of core-crust model of quaking neutron star (Franco et al 2000),  there are two way of thinking why only peripheral region of the star can be set in Lorentz-force-driven Alfv\'en vibrations.  First belong to a case when strength of seismic tremor is fairly week and perturbation sets in vibrations only finite-depth crustal region of the star, leaving massive core unaltered. The second way, when core material
  just incapable of sustaining Alfv\'en vibrations. One of the main conditions for existing of such vibrations
  is the extremely large (effectively infinite) electrical conductivity of stellar substance (e.g., Chandrasekhar 1961, Mestel 1999), but this may not be the case for core material whose microscopic composition
  is dominated, according to canonical understanding of neutron star, by matter possessing properties of
  non-conducting degenerate Fermi-gas of non-relativistic neutrons.

  \begin{figure}
\centering{\includegraphics[width=8.0cm]{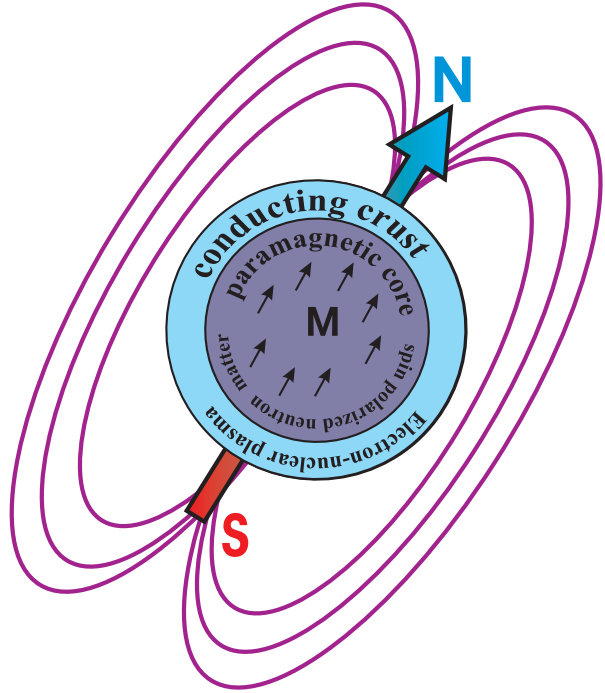}}
\caption{
 The internal constitution of two-component, core-crust, model of paramagnetic neutron star.
 The massive core is considered as a poorly conducting permanent magnet composed of degenerate Fermi-gas of non-relativistic neutrons in the state of Pauli's paramagnetic saturation caused by field-induced alignment of spin magnetic moment of neutrons along the axis of uniform internal and dipolar external magnetic field frozen in the star on the stage of gravitational collapse of its MS progenitor. A highly conducting metal-like material of the neutron star crust, composed of nuclei embedded in the super dense degenerate Fermi-gas  of relativistic electrons, is regarded as  electron-nuclear solid-state magneto-active plasma capable of sustaining Alfv\'en oscillations.}
\end{figure}

  The physically meaningful framework for discussion of this second case provides the two-component, core-crust,
  model of quaking paramagnetic neutron star, pictured in Fig.1, that has been extensively discussed some time ago
  in works \footnote{The main purpose of these works was to explore
  potential capabilities of equations of magneto-elastic dynamics (MED) suggested as fundamental equations of continuum mechanics and macroscopic electrodynamics of non-conducting but highly magnetically polarized stellar matter
  that have been formulated in a manner of equations of magnetohydrodynamics underlying continuum-mechanical
  treatment of motions of perfectly conducting liquid-state and solid-state plasmas pervaded by magnetic fields.
  It was found on the basis of MED equations that magnetically polarized paramagnetic neutron matter, with spin-magnetic moments of neutrons aligned along frozen-in magnetic field, can transmit perturbation by torsional magneto-elastic wave having for such a matter one and the same physical meaning as Alfv\'en wave does for
  magneto-active liquid-state or solid-state plasmas. Also, it was shown that a neutron star whose material
  is in the state of paramagnetic permanent magnetization can sustain solely torsional oscillations
  driven by force of antisymmetric magneto-polarizational stresses.} (Bastrukov et al 2001, 2002a, 2002b, 2003).
 In this model, the role of major source of super strong magnetic field of the star is attributed to
 permanently magnetized massive core composed of degenerate neutron matter in the the state
 of Pauli's paramagnetic saturation in which magnetic moments of neutrons turn out aligned in the direction of
 frozen-in magnetic field that has been inherited by neutron star from massive progenitor and amplified in the implosive process of its gravitational collapse. Unlike the non-conducting core matter, the material
 of crust matter resembles metallic solid. This suggests that it is magnetic cohesion between metal-like crust and permanent-magnet-like core, mediated by lines of ultra strong magnetic field, should play main part in the
 the starquake dynamics. In this paper we adopt this latter point of view by considering starquake
 as impulsive release of magnetic field energy by means of disruption of magnetic field lines on the core-crust interface and fracturing the crust by relived magnetic stresses.
 In some sense the magnetic field lines can be thought of as a super-hard piles endowing core-crust composition of neutron star with supplementary (to gravity forces) stiffness which is primarily responsible for its seismic stability to quake-induced tectonic shear displacements of crust against core\footnote{The internal constitution of paramagnetic neutron star model under consideration does not exclude the existence of a fairly thin inner crust
 with average density several times less than the normal nuclear density (neutron drip-line)
 at which micro-composition of matter is dominated, as we know from nuclear physics, by paired neutrons. It must be stated clear, however, that such quasi-boson matter composed of paired neutrons has nothing to do with fairly dilute and flowing quasi-boson matter like superfluid Fermi-liquid composed of $^3$He atoms. According to current theory, the emergence of superfluidity in flowing boson material substances belongs to phase transition of macroscopic fraction of bosons to the state of Bose-Einstein condensation (BEC). One of the most striking features of such phase transition is that at the temperature of condensation the internal pressure of BEC state vanishes.
 The fact that Bose-matter composed of paired neutrons in BEC state is unable to withstand the pressure of self-gravity (contrary to the degenerate neutron Fermi-matter whose zero-temperature pressure of degeneracy counterbalancing the self-gravity pressure is central to the very notion of neutron star) suggests that such a matter, if exists, can only insufficiently contribute to the total mass budget of neutron star, otherwise the very notion of this compact object is loosing its original meaning. In the starquake of paramagnetic neutron star model under consideration,
 this inner crust material is thought of as operating like a lubricant facilitating
  non-compressional differentially rotational shear displacements of crust relative to much denser matter of massive core.}. It is expected that quake-induced disruptions of magnetic field lines, along which the outgoing radiation is developed, should be observed as glitches in the quiescent high-energy emission. These glitches
  are of substantially magnetic nature in the sense that they owe its origin to well-known effect of sudden, jump-like, increase in the magnetic field strength triggered by thermonuclear explosion.
  With this in mind we focus on reaction of the star on above quake-induced perturbation
  by node-free differentially rotational, torsional, vibrations of crustal solid-state plasma about axis of magnetic field frozen in the immobile paramagnetic core.

  In section 2, a brief outline is given of the equations of solid-magnetics
  appropriate for the perfectly conducting viscoelastic continuous medium pervaded by a magnetic field.
  In section 3, the spectral formulae for the frequency and lifetime of this toroidal mode of non-radial nodeless vibrations driven by combined magnetic Lorentz and elastic Hooke's forces are obtained.
  The relevance of considered asteroseismic model to the detected QPOs in the X-ray flux
  during the flare of SGR 1806-20 and SGR 1900+14 is assessed in section 4. The newly obtained results
  are briefly summarized in section 5.

\section{Governing equations of solid-magnetics}

  We study an asteroseismic model of non-rotating neutron star responding
  to quake-induced perturbations by shear (substantially non-compressional, $\delta\rho=-\rho\nabla_k u_k=0$) differentially rotational fluctuations of material displacements $u_i$, locked in the finite-depth
  crust of the density $\rho$. And it is believed that dynamics of this vibrational process can be
  properly modeled by equations of classical solid mechanics
  \begin{eqnarray}
  \label{e2.1}
  \rho{\ddot u}_i=\nabla_k\,\tau_{ik}+\nabla_k\,\sigma_{ik}+\nabla_k\,\pi_{ik},\quad \nabla_k u_k=0
 \end{eqnarray}
  presuming that Hooke's elastic stresses $\sigma_{ik}$ and Newton's viscous stresses $\pi_{ik}$ are described
  by linear constitutive equations of the form
  \begin{eqnarray}
 \label{e2.2}
 && \sigma_{ik}=2\mu\,{u}_{ik},\quad {u}_{ik}=\frac{1}{2}[\nabla_i {u}_k+\nabla_k {u}_i],\\
 && \label{e2.3}
 \pi_{ik}=2\eta{\dot u}_{ik},\quad {\dot u}_{ik}=\frac{1}{2}[\nabla_i {\dot u}_k+\nabla_k {\dot u}_i]
 \end{eqnarray}
  where $\mu$ stands for the shear modulus, $\eta$ for shear viscosity and $u_{ik}$ is the tensor of shear strains or deformations. The central to our further discussion is the tensor of fluctuating magnetic field stresses
  \begin{eqnarray}
 \label{e2.4}
 &&\tau_{ik}=\frac{1}{4\pi}[B_i\delta B_k+B_k\delta B_i -B_j\delta B_j \delta_{ik}],\\
 \label{e2.4a}
 &&\delta B_i=\nabla_k\,[u_i\,B_k-u_k\,B_i]
 \end{eqnarray}
 As in our previous work (Bastrukov et al 2009a), we consider model with homogeneous internal
 magnetic field whose components in spherical polar coordinates read
  \begin{eqnarray}
  \label{e2.5}
   && B_r=B\cos\theta,\quad\quad B_\theta=-B\sin\theta,
 \quad\quad B_\phi=0
 \end{eqnarray}
  and external dipolar magnetic field is described by ${\bf B}=\nabla\times {\bf A}$, where
  ${\bf A}=[0, 0, A_\phi={\rm m}_s/r^2]$ is the vector potential with the standard parametrization of the dipole magnetic moment ${\rm m}_s=(1/2)BR^3$ of star of radius $R$ and by $B$ is
  understood the magnetic field intensity at its magnetic poles, $B=B_p$.

\subsection{The energy method}

 This method of computing frequency of shear vibrations rests on the equation of energy balance
 \begin{eqnarray}
  \label{e2.6}
 \frac{\partial }{\partial t} \int \frac{\rho {\dot u}^2}{2}\,
 d{\cal V}=-\int [\tau_{ik}+\sigma_{ik}+\pi_{ik}]\,{\dot u}_{ik}\,d{\cal V}
 \end{eqnarray}
 which is obtained by scalar multiplication of equation of magneto-solid-mechanics (\ref{e2.1}) with $u_i$ and integration over the volume of seismogenic layer. From the technical point of view,
 the shear character of material distortions brought about by forces under consideration owes its origin to
 the symmetric form of stress-tensors in terms of which these forces are expressed.
 The key idea of this method consists in using of the following separable form of material displacements
 \begin{eqnarray}
 \label{e2.7}
 && u_i({\bf r},t) =a_i({\bf r}){\alpha}(t)
 \end{eqnarray}
 where $a_i({\bf r})$ is the time-independent solenoidal field and amplitude ${\alpha}(t)$
 carries information about temporal evolution of fluctuations. Thanks to this latter substitution, the
 all tensors of fluctuating stresses and strains take similar separable form
 \begin{eqnarray}
  \label{e2.8}
 && \tau_{ik}({\bf r},t)=[{\tilde \tau}_{ik}({\bf r})-\frac{1}{2}{\tilde \tau}_{jj}({\bf r})\delta_{ik}]\alpha(t),\\
  \label{e2.9}
 && {\tilde \tau}_{ik}({\bf r})=\frac{1}{4\pi}
 [B_i({\bf r})\,b_k({\bf r})+B_k({\bf r}),\,\, b_i({\bf r})],\\
 \label{e2.10}
 && b_i({\bf r})=\nabla_k\,[a_i({\bf r})\,B_k({\bf r})-a_k({\bf r})\,B_i({\bf r})],\\
 \label{e2.11}
 && {\sigma}_{ik}({\bf r})=2\mu\,a_{ik}({\bf r})\alpha(t),\quad {\pi}_{ik}({\bf r})=2\eta\,a_{ik}({\bf r})\alpha(t),\\
  \label{e2.12}
 && a_{ik}({\bf r})=\frac{1}{2}[\nabla_i a_k({\bf r})+\nabla_k a_i({\bf r})].
 \end{eqnarray}
 On inserting (\ref{e2.7})-(\ref{e2.11}) in the integral equation of energy balance (\ref{e2.6})
 we arrive at equation for $\alpha(t)$ having the well-familiar form
\begin{eqnarray}
 \label{e2.13}
 && \frac{d{\cal E}}{dt}=-2{\cal F},\quad {\cal E}=\frac{{\cal M}{\dot \alpha^2}}{2}+\frac{{\cal K}\alpha^2}{2},\quad
 {\cal F}=\frac{{\cal D}{\dot \alpha^2}}{2}\\
 \label{e2.14}
 &&{\cal M}{\ddot \alpha}+{\cal D}{\dot
 \alpha}+{\cal K}\alpha=0,\\
  \label{e2.14a}
 &&\alpha(t)=\alpha_0\exp(- t/\tau)\cos(\Omega t),\\
 \label{e2.15}
 && \quad \Omega^2=\omega^2\left[1-(\omega\tau)^{-2}\right],\quad
  \omega^2=\frac{{\cal K}}{{\cal M}},\quad \tau=\frac{2{{\cal M}}}{{\cal D}}.
 \end{eqnarray}
 where  the inertia ${\cal  M}$, viscous friction ${\cal  D}$ and stiffness ${\cal  K}$
 of oscillator are given by
 \begin{eqnarray}
  \label{e2.16}
  && {\cal M}=\int \rho({\bf r}) a_i({\bf r})\,a_i({\bf r})\,d{\cal V},\quad {\cal K}={\cal K}_e+{\cal K}_m,\\
 \label{e2.17}
  && {\cal K}_e= 2\int \mu(r)\,a_{ik}({\bf r})\,a_{ik}({\bf r})\,d{\cal V},\\
 \label{e2.18}
 && {\cal K}_m=\int \,{\tilde \tau}_{ik}({\bf r})\,a_{ik}({\bf r})\,d{\cal V},\\
 \label{e2.19}
  && {\cal D}= 2\int \eta(r) \,a_{ik}({\bf r})\,a_{ik}({\bf r})\,d{\cal V}.
 \end{eqnarray}
 The physical significance and practical usefulness of the outlined energy method
 is that it can be efficiently utilized in the study of non-radial seismic vibrations of more wide class of
 solid degenerate stars like white dwarfs stars (Lou 1995, Molodtsova et al 2010) and ultra dense quark-matter stars\footnote{The average density of quark star matter is expected to be comparable with the average density of nucleon $\rho\approx 4.3\times 10^{17}$ g cm$^{-3}$, that is, three order of magnitude greater than average density of normal nuclear matter $\rho\approx 2.8\times 10^{14}$ g cm$^{-3}$.} whose material, as is expected
 (Xu 2009 and references therein), can too be thought of as an elastic solid extremely robust to compressional distortions. In (Bisnovatyi-Kogan 2008, Heyvaerts et al 2009) it is proposed that ultra powerful luminosity of currently monitored $\gamma$-ray sources could be related to torsional vibrations of neutron and quark stars.

\subsection{Material displacements in the crust undergoing node-free torsional vibrations}

  In this subsection we pause on detailed derivation of material displacements in the crust
  undergoing non-radial torsional oscillations in the node-free regime.
  In this mode,  the velocity of fluctuating material flow (the rate of material displacements)
  is described by general formula of rotational motions
  \begin{eqnarray}
  \label{e3.1}
  && \delta {\bf v}({\bf r},t)={\dot {\bf u}}({\bf r},t)=[\mbox{\boldmath $\Omega$}({\bf r},t)\times {\bf r}],\\
  \label{e3.2}
  && \mbox{\boldmath $\Omega$}({\bf r},t)=
  [\nabla\times \delta {\bf v}({\bf r},t)]=[\nabla\times {\dot {\bf u}}({\bf r},t)]={\dot{ \mbox{\boldmath $\Phi$}}}({\bf r},t)
  \end{eqnarray}
   However, unlike a case of rigid-body rotation, in which the angular velocity is a constant vector,
   in a solid mass undergoing axisymmetric differentially rotational vibrations the angular velocity $\mbox{\boldmath $\Omega$}({\bf r},t)$ is the vector-function of position (Bastrukov et al 1999a, 2007a) which
   can be represented as
 \begin{eqnarray}
  \label{e3.3}
  && \mbox{\boldmath $\Omega$}({\bf r},t)={\dot{ \mbox{\boldmath $\Phi$}}}({\bf r},t)=\mbox{\boldmath $\phi$}({\bf r})\,{\dot \alpha}(t),\quad\quad \mbox{\boldmath $\phi$}({\bf r})=[\nabla\times {\bf a}({\bf r})]
  \end{eqnarray}
  In the regime of node-free vibrations in question, ${\bf a}({\bf r})$ is described by the divergence-free odd-parity, axial, toroidal field which is one of two harmonic solenoidal fields of fundamental basis (Chandrasekhar 1961)
  obeying the vector Laplace equation $\nabla^2 {\bf a}=0$. This field can be expressed
  in terms of general solution of the scalar Laplace equation as follows
   \begin{eqnarray}
 \label{e3.4}
 && {\bf a}({\bf r})={\bf a}_t({\bf r})=\nabla \times [{\bf r}\,\chi({\bf r})]=
 [\nabla \chi({\bf r}) \times {\bf r}]\\
  \label{e3.5}
 && \nabla^2\chi({\bf r})=0,
 \\
 && \chi({\bf r})=[A_\ell\,r^\ell+B_\ell\,r^{-\ell-1}]\,P_\ell(\zeta),\quad \zeta=\cos\theta
  \end{eqnarray}
 where $P_\ell(\zeta)$ is the Legendre polynomial of multipole degree $\ell$. It follows that the angular field $\mbox{\boldmath $\phi$}({\bf r})$ is the poloidal vector field
  \begin{eqnarray}
 \label{e3.6}
 \mbox{\boldmath $\phi$}({\bf r})&=&[\nabla\times {\bf a}_t({\bf r})]=\nabla\times \nabla\times [{\bf r}\,\chi({\bf r})]\\
 &=&\nabla\,[{\cal A}_\ell\,r^\ell+{\cal B}_\ell\,r^{-\ell-1}]\,P_\ell(\zeta),\\
 && {\cal A}_\ell=A_\ell(\ell+1),\quad {\cal B}_\ell=B_\ell\,\ell.
 \end{eqnarray}
 And this field is irrotational, $\nabla\times \mbox{\boldmath $\phi$}({\bf r})=0$.

\begin{figure}[h]
\centering{\includegraphics[width=7.5cm]{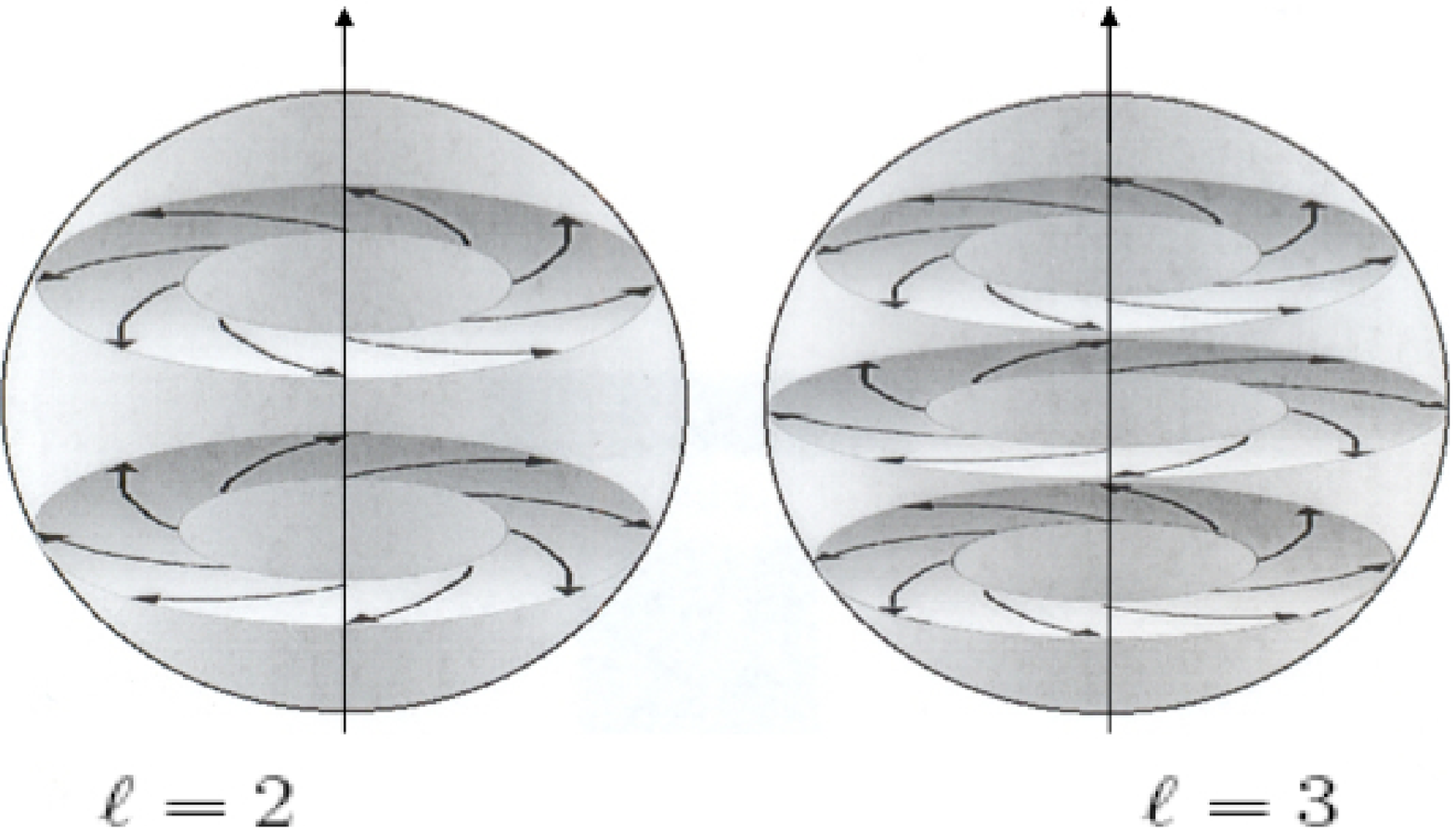}}
 \caption{Material displacements of crustal matter about the dipole magnetic moment axis of paramagnetic neutron star undergoing nodeless differentially rotational, torsional, vibrations in quadrupole and octupole overtones.}
\end{figure}

 As was stated, we study a model of differentially rotational vibrations of peripheral finite-depth crust against
 immobile core.  In this case, the arbitrary constants ${\cal A}_\ell$ and ${\cal B}_\ell$ can be
 uniquely eliminated from two boundary conditions: (i) on the core-crust interface $r=R_c$
 \begin{eqnarray}
   \label{e3.7}
 u_\phi\vert_{r=R_c}=0
 \end{eqnarray}
and (ii) on the star surface $r=R$
  \begin{eqnarray}
   \label{e3.8}
 && u_{\phi}\vert_{r=R}=[\mbox{\boldmath $\Phi$}\times {\bf R}]_\phi\vert_{r=R},
 \\
 && \mbox{\boldmath $\Phi$}=\alpha(t)\nabla_{\hat n} P_\ell(\zeta),\quad {\bf R}={\bf e}_rR
 \end{eqnarray}
 where
 \begin{eqnarray}
 \label{e3.9}
  \nabla_{\hat n}=\frac{1}{R}\nabla_{\mbox{\boldmath $\Omega$}},\quad \nabla_{\mbox{\boldmath $\Omega$}}= \left[{\bf e}_\theta \frac{\partial }{\partial
 \theta}+{\bf e}_\phi\frac{1}{\sin\theta}\frac{\partial }{\partial \phi}\right].
 \end{eqnarray}
 The no-slip condition on the core-crust interface, $r=R_c$,
 reflects the fact that the amplitude of differentially rotational oscillations
 gradually decreases down to the star center and turns into zero on the core.
 The boundary condition on the  star surface, $r=R$,
 is dictated by symmetry of the general toroidal solution of the vector Laplace equation which then
 is tested to reproduce the moment of inertia of a rigidly rotating solid star
 (Bastrukov et al 2008a). The above boundary conditions lead to the coupled algebraic equations
 \begin{eqnarray}
 \label{e3.10}
&&{\cal A}_\ell R_c^{\ell-1}+{\cal B}_\ell R_c^{-\ell-2}=0,\quad\quad {\cal A}_\ell R^{\ell}+{\cal B}_\ell R^{-\ell-1}=R
\end{eqnarray}
whose solutions are
\begin{eqnarray}
  \label{e3.11}
 {\cal A}_\ell={\cal N}_\ell,\quad {\cal B}_{\ell}=-{\cal
 N}_\ell\,R_c^{2\ell+1},\quad\quad {\cal
 N}_\ell=\frac{R^{\ell+2}}{R^{2\ell+1}-R_c^{2\ell+1}}.
 \end{eqnarray}
In spherical polar coordinates, the nodeless toroidal field has only one non-zero azimuthal component
\begin{eqnarray}
 \label{e3.12}
 && a_{r}=0,\,\, a_{\theta}=0,\\
 && a_{\phi}=\left[{\cal A}_\ell\,r^\ell+\frac{{\cal B}_\ell}{r^{\ell+1}}\right](1-\zeta^2)^{1/2}\frac{d P_\ell(\zeta)}{d\zeta}.
 \end{eqnarray}
 The snapshot of material node-free displacements in the crust undergoing torsional oscillations against immobile core of paramagnetic neutron star under consideration is pictured in Fig.2 for
 quadrupole, $\ell=2$, and octupole $\ell=3$, overtones of this axial mode.

\section{Spectral formulae for the frequency and lifetime of seismic vibrations}

 The computation of integrals defining mass parameter ${\cal M}$, parameter of vibrational rigidity ${\cal K}$ and viscous friction ${\cal D}$ [which has been presented in some details elsewhere (Bastrukov et all 1999, 2007a, 2009)] are quite lengthy but straightforward and, therefore,
 are not presented here. The mass parameter can be conveniently
 represented in the form
\begin{eqnarray}
 \label{e4.1}
 && {\cal M}=4\pi\,\rho R^5\frac{\ell(\ell+1)}{(2\ell+1)(2\ell+3)}\,m(\ell),\\
 \label{e4.1a}
 && m(\ell)=(1-\lambda^{2\ell+1})^{-2}\\
 \nonumber
 && \times \left[1-
 (2\ell+3)\lambda^{2\ell+1}
  \right. +\\ \nonumber
 && +  \left.
 \frac{(2\ell+1)^2}{2\ell-1}\lambda^{2\ell+3}-
 \frac{2\ell+3}{2\ell-1}\lambda^{2(2\ell+1)}\right],\\
 \label{e4.2}
 && \lambda=\frac{R_c}{R}=1-h,\quad h=\frac{\Delta R}{R},\\
 \label{e4.2a}
 && \Delta R=R-R_c,\quad 0\leq \lambda <1.
\end{eqnarray}
The $\lambda$-terms in the above and foregoing equations emerge as result of integration
along the radial coordinate from radius of the core-crust interface $r=R_c$ to the star radius, $r=R$.
The integral coefficient of viscous friction is given by
\begin{eqnarray}
\label{e4.3}
 && {\cal D}=4\pi\eta R^3\frac{\ell(\ell^2-1)}{2\ell+1}
 \,d(\ell),\\
 \label{e4.4}
 && d(\ell)={(1-\lambda^{2\ell+1})^{-1}}
 \left[1-\frac{(\ell+2)}{(\ell-1)}\lambda^{2\ell+1}\right].
\end{eqnarray}

 \begin{figure}[h]
\centering{\includegraphics[width=8.cm]{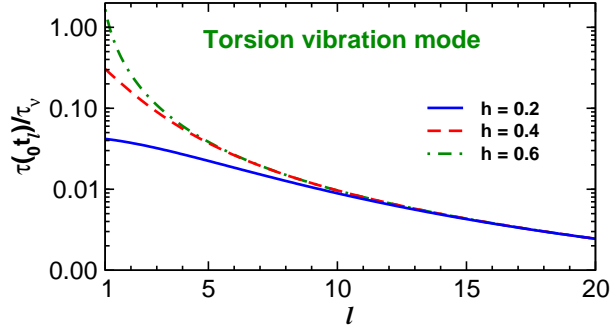}}
 \caption{The fractional lifetime of torsion nodeless oscillations of the neutron star crust damped by force of viscous shear stresses as a function of multipole degree $\ell$ computed at indicated values of the fractional depth $h$ of peripheral seismogenic layer.}
\end{figure}

For the lifetime we obtain
 \begin{eqnarray}
 \label{e4.5}
 && \tau(_0t_\ell)=\frac{2\tau_\nu}{(2\ell+3)(\ell-1)}\,\frac{m(\ell)}{d(\ell)},
 \quad \tau_\nu=\frac{R^2}{\nu},\quad \nu=\frac{\eta}{\rho}.
\end{eqnarray}
 In Fig.3, the fractional lifetime
 is plotted as a function of multipole degree $\ell$ with indicated values of fractional depths $h=\Delta R/R$.
 It shows that the higher $\ell$ the shorter lifetime. It is easy to see that in the limit, $\lambda=(R_c/R)\to 0$, we regain the spectral formula for lifetime of global torsional nodeless vibrations of solid star (Bastrukov et al 2003, 2007a)
\begin{eqnarray}
  \label{e4.6}
 && \tau(_0t_\ell)=\frac{2{\bar\tau}_\nu}{(2\ell+3)(\ell-1)},\quad {\bar \tau}_\nu=\frac{R^2}{\bar\nu},\quad {\bar \nu}=\frac{\bar\eta}{\bar\rho}.
 \end{eqnarray}
 in which by $\bar\tau_\nu$ is understood, in this latter case,  the average kinematic viscosity
 of the star matter as a whole; the extensive discussion of this transport coefficient can be found in (Shternin 2008). For the node-free torsional oscillations of solid star, the last equation
 has one and the same physical significance as the well-known Lamb formula does for the
 time of viscous damping of spheroidal node-free vibrations which
 in the context of neutron star pulsations has been extensively discussed some time ago by Cutler and Lindblom (1987).
 Regarding the problem under consideration we cannot see, however, how
 the obtained formulae can be applied to observational data on QPOs in SGRs.
 Nonetheless, their practical usefulness is that they can be utilized in the study
 of a more wide class of celestial objects such as Earth-like planets (Lay \& Wallace 1995, Aki \& Richards 2003) and white dwarf stars.

 From above it is clear that the integral coefficient of elastic rigidity ${\cal K}_e$ of torsional vibrations
 has analytic form similar to that for coefficient of viscous friction ${\cal D}$, namely
\begin{eqnarray}
\label{e4.7}
 && {\cal K}_e=4\pi\mu R^3\frac{\ell(\ell^2-1)}{2\ell+1}
 \,k_e(\ell),\\
\label{e4.8}
 && k_e(\ell)={(1-\lambda^{2\ell+1})^{-1}}
 \left[1-\frac{(\ell+2)}{(\ell-1)}\lambda^{2\ell+1}\right].
\end{eqnarray}
The frequency as a function of multipole degree  $\ell$  of node-free elastic vibrations in question $\nu_e(\ell)$ (measured in Hz and related to angular frequency as $\omega_e(\ell)=2\pi\nu_e(\ell)={\cal K}_e/{\cal M}$) is given by
 \begin{eqnarray}
 \label{e4.9}
 && \nu_e^2(\ell)=\nu^2_e\left[(2\ell+3)(\ell-1)\right]^{1/2}\,
 \frac{k_e(\ell)}{m(\ell)},\\
 \label{e4.10}
 && \omega_e=2\pi\nu_e=\frac{c_t}{R},\quad c_t=\sqrt{\frac{\mu}{\rho}},\quad
 \lambda=1-h,\,\, h=\frac{\Delta R}{R}
 \end{eqnarray}
 It is easy to see that in the limit $\lambda\to 0$,  we regain spectral formula for the frequency of global torsional oscillations, having the form of equation (\ref{e4.9}) with $k_e(\ell)=m(\ell)=1$. Understandably that in this latter case all material characteristics belong to the star as a whole.

 The magneto-mechanical stiffness of Alfv\'en vibrations ${\cal K}_m$
 can conveniently be written as
\begin{eqnarray}
 \label{e4.11}
 && K_m=B^2R^3\frac{\ell(\ell^2-1)(\ell+1)}{(2\ell+1)(2\ell-1)}\,k_m(\ell)\\
 \label{e4.12}
 && k_m(\ell)=(1-\lambda^{2\ell+1})^{-2}\times \left\{1+\frac{3\lambda^{2\ell+1}}{(\ell^2-1)(2\ell+3)}\
  \right. \times \\ \nonumber
 && \times \left.
 \left[1-\frac{1}{3}\ell(\ell+2)(2\ell-1)
 \lambda^{2\ell+1}\right]\right\}.
 \end{eqnarray}
 This leads to the following two-parametric spectral formula
 \begin{eqnarray}
 \label{e4.13}
 && \nu_m^2(\ell)=\nu^2_A\left[(\ell^2-1)\frac{2\ell+3}{2\ell-1}\right]\,
 \frac{k_m(\ell)}{m(\ell)},\\
 \label{e4.14}
 && \omega_A=2\pi\nu_A=\frac{v_A}{R},\quad v_A=\frac{B}{\sqrt{4\pi\rho}}
 \end{eqnarray}
 In the forward asteroseismic analysis of QPO data
 relying with on this latter spectral formula, the  Alfv\'en frequency, $\nu_A$ and the fractional depth of seismogenic zone, $h$, are regarded as free parameters which are adjusted so as to reproduce general trends in the observed
 QPOs frequencies.

\begin{figure}[h]
\centering{\includegraphics[width=8.cm]{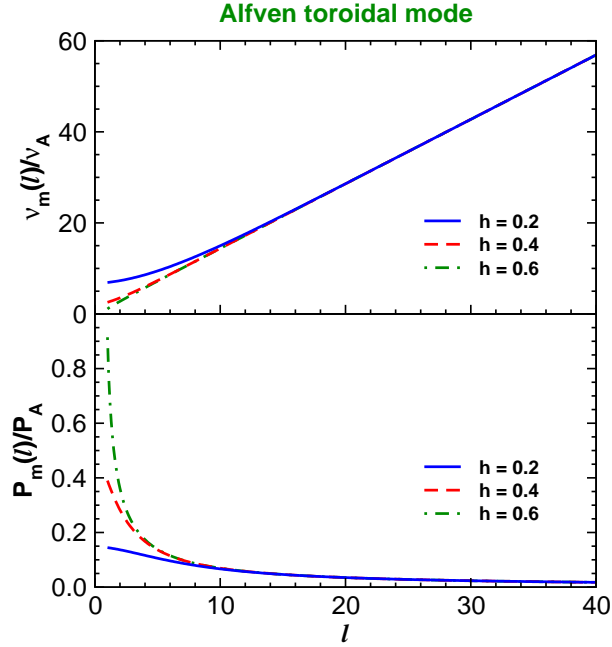}}
 \caption{Fractional frequency and period of nodeless torsional magneto-solid-mechanical oscillations,
 toroidal Alfv\'en mode -- $_0a^t_\ell$, entrapped in the neutron star crust as functions of multipole degree $\ell$
 computed at indicated values of the fractional depth $h$ of peripheral seismogenic layer. The value
 $h=1$ corresponds to global torsional oscillations excited in the entire volume of the star. Here $\nu_A=\omega_A/2\pi$, where $\omega_A=v_A/R$ with $v_A=B/\sqrt{4\pi\rho}$ being the velocity
 of Alfv\'en wave in crustal matter of density $\rho$ and $P_A=2\pi/\omega_A$.}
\end{figure}

  In Fig.4, the fractional frequencies and periods of this toroidal Alfv\'en mode
  as functions of multipole degree $\ell$
 are plotted with indicated values of fractional depth of the seismogenic layer $h$.
 Remarkably, the lowest overtone of global oscillations is of quadrupole degree, $\ell=2$, whereas for
 vibrations locked in the crust, the lowest overtone is of dipole degree, $\ell=1$, as is clearly seen
 in Fig.5. This suggests that dipole vibration can be thought of as
 Goldstone's soft mode whose most conspicuous property is that the mode
 disappears (the frequency tends to zero) when key parameter regulating the depth of seismogenic zone $\lambda\to 0$.

\begin{figure}[h]
\centering{\includegraphics[width=8.cm]{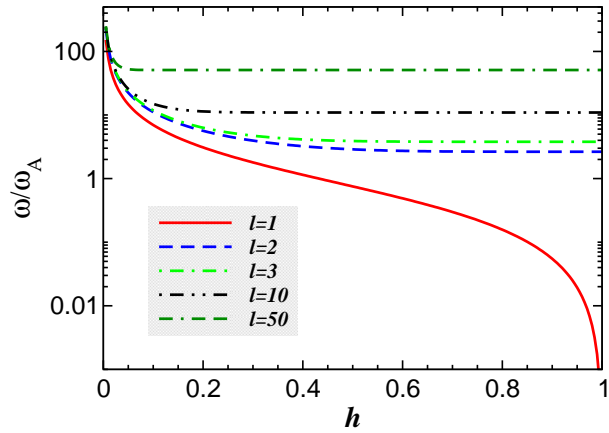}}
 \caption{Fractional frequency of nodeless torsional Alfv\'en oscillations
 of indicated overtones $\ell$ as a function of the fractional depth $h$ of peripheral seismogenic layer.
 The vanishing of dipole overtone in the limit of $h\to 1$, the case when entire mass of neutron star sets in torsional oscillations, suggests that dipole vibration possesses property typical to Goldstone's soft modes .}
\end{figure}

\section{Application to QPOs in the outburst X-ray flux from SGR 1806-20 and SGR 1900+14}

  The basic physics underlying current understanding of interconnection between quasi-periodic oscillations of detected electromagnetic flux and vibrations of neutron star has been recognized long ago (e.g. Boriakoff 1976, van Horn 1980, Blaes et al 1988). Owing to the effect of strong flow-field coupling, which is central to
  the propagation of Alfv\'en waves, the quake induced perturbation excites coupled vibrations of perfectly conducting solid-state plasma of the crust (as well as gaseous plasma of magnetar corona expelled from the surface by outburst) and frozen-in lines of magnetic field. Outside the star the vibrations
  of magnetic field lines are coupled with oscillations of gas-dust plasma expelled from the star surface by
  quake. And it is these fluctuations of outer lines of magnetic field, operating like transmitters of beams of charged particles producing coherent (curvature and/or synchrotron) high-energy radiation, are detected as QPOs
  of  light curves of the SGRs giant flares.

  In applying the obtained spectral formulae to the frequencies of detected QPOs we examine two scenarios, namely, when quake-induced torsional vibrations are restored by joint action of Lorentz magnetic and Hooke's
  elastic forces and when oscillations are of pure Alfv\'en's nature, that is, produced by torsional seismic
  vibrations of crust against core under the action of solely one Lorentz force of magnetic field stresses.

\subsection{Crust vibrations driven by combined action of Loternz magnetic and Hooke's elastic forces}

In this case, the asteroseismic analysis of detected QPOs rests on the three-parametric spectral formula
   \begin{eqnarray}
 \label{e5.1}
 && \nu^2(\ell)[\nu_A,\nu_e,h]=\nu_m^2(\ell)[\nu_A,h]+\nu^2_e(\ell)[\nu_e,h]
 \end{eqnarray}
  The suggested theoretical $\ell$-pole specification of the detected frequencies
  is presented in Fig. 6 for SGR 1900+14 and in Fig. 7 and Fig. 8, exhibiting remarkable correlation between depth  of seismogenic zone the fundamental frequencies of coherent magnetic and elastic oscillations - the larger $\Delta R$, the higher basic frequencies of Alfv\'enic $\nu_A$ and elastic $\nu_e$ vibrations. It is seen from computations
  for SGR 1806-20, that  reasonable fit of data can be attained with $h=0.2$ (for the star model with radius 20 km, $\Delta R=2$ km) and with $h=0.4$ ($\Delta R=5$ km).

   \begin{figure}[h]
 \centering{\includegraphics[width=8.0cm]{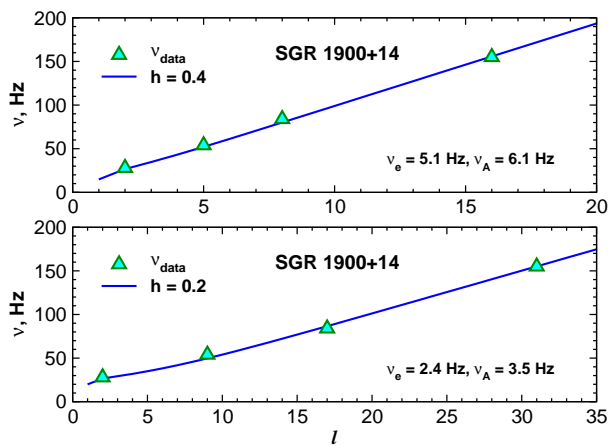}}
 \caption{Theoretical fit of the QPOs frequency in the X-ray flux from  SGR 1900+14 on the basis
 of three-parametric theoretical spectrum of frequency of torsional seismic
 vibrations in the crustal region of indicated fractional depth.}
\end{figure}

    \begin{figure}[h]
 \centering{\includegraphics[width=8.0cm]{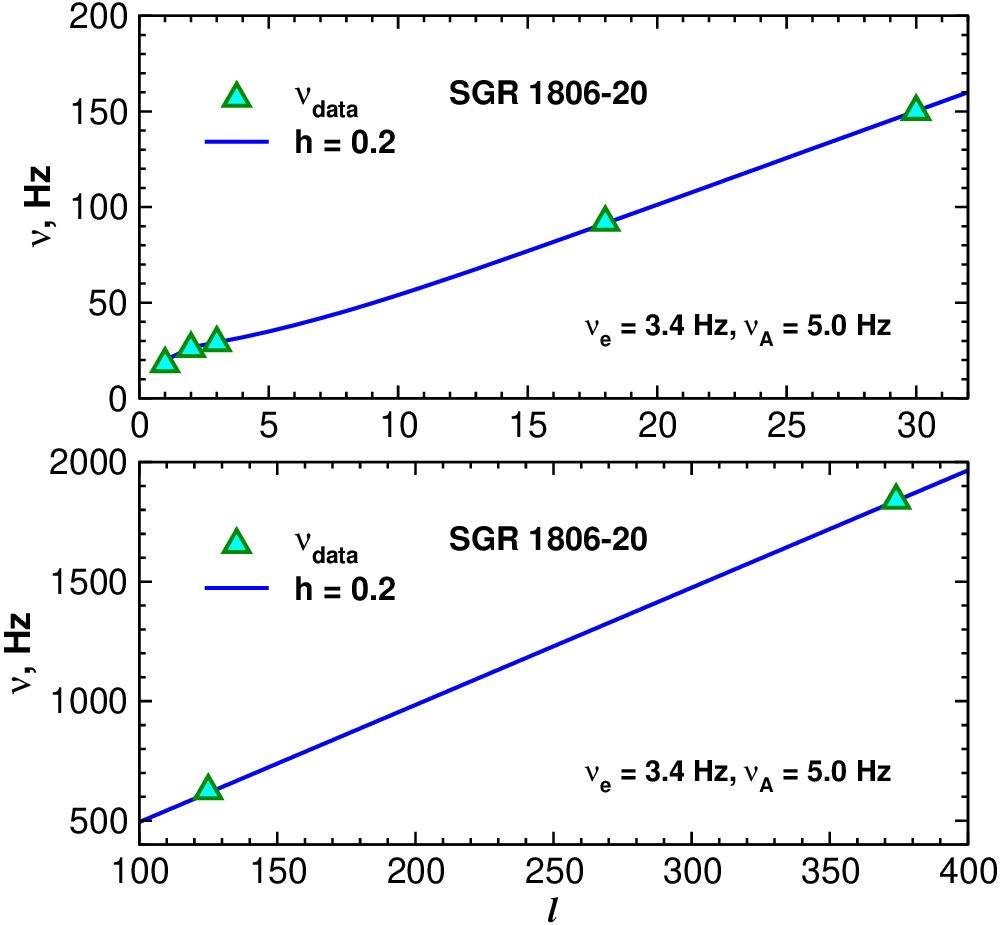}}
 \caption{The same as Fig.6, but for SGRs 1806-20 with h=0.2.}
\end{figure}

 \begin{figure}[h]
 \centering{\includegraphics[width=8.0cm]{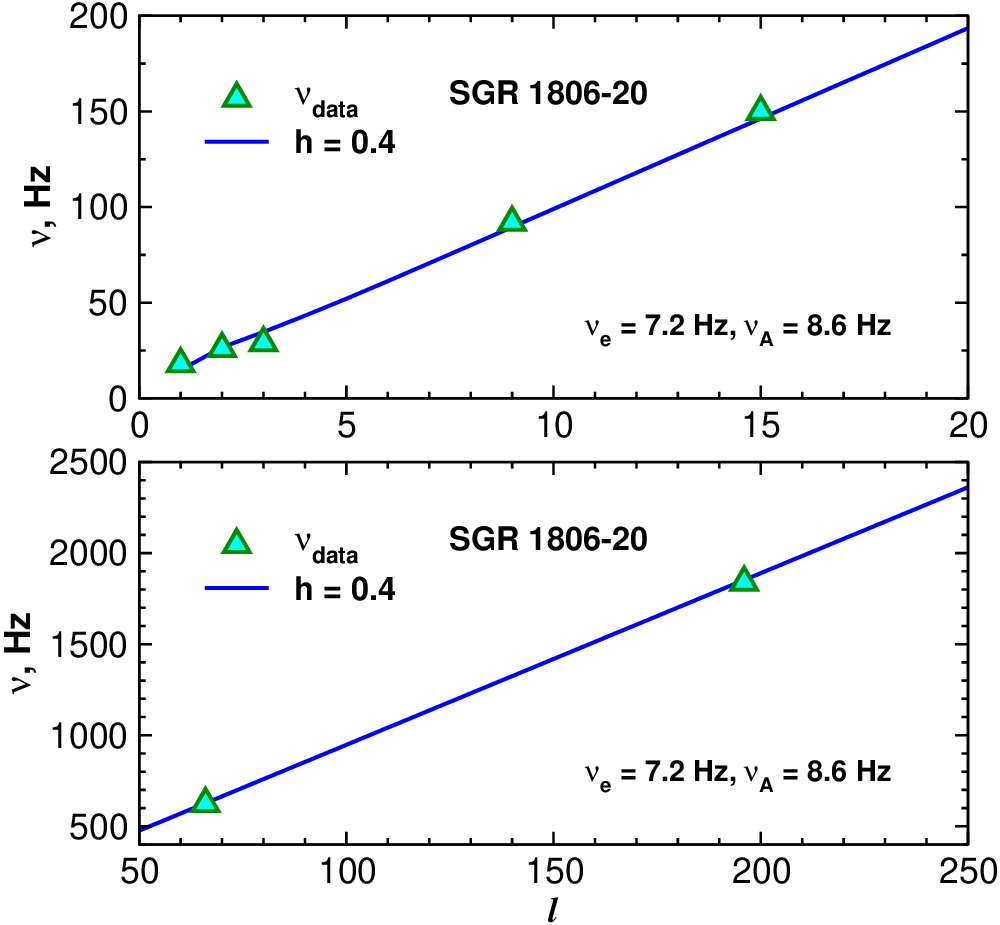}}
 \caption{The same as Fig.6, but for  SGRs 1806-20 with h=0.4.}
\end{figure}

  It is worth emphasizing that at above values of $h$, the obtained here tree-parametric spectral formula much better match the data as compared to that for global, in the entire volume, vibrations studied in (Bastrukov et al 2009a).
  On this ground we conclude, if the detected QPOs are produced by seismic vibrations of peripheral region of the star under coherent action of Lorentz and Hooke's forces, then the depth of seismogenic layer $\Delta R$ should be quite large, somewhere in the range $0.2R < \Delta R < 0.4R$.

\subsection{Seismic vibrations of peripheral layer under the action of Loternz magnetic force}

 In this case $\ell$-pole specification of detected QPOs frequencies rests on
 the two-parametric spectral formula for frequency of purely Alfv\'en toroidal mode
 \begin{eqnarray}
 \label{e5.2}
 && \nu^2(\ell)=\nu_m^2(\ell[v_A,h]).
 \end{eqnarray}

  \begin{figure}[h]
 \centering{\includegraphics[width=8.0cm]{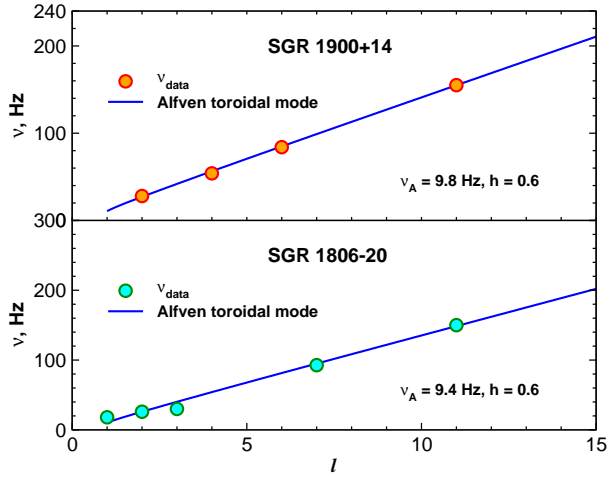}}
 \caption{Theoretical description (lines) of detected QPO frequencies (symbols) in the X-ray flux during the flare of SGRs 1806-20 and 1900+14 as overtones of pure Alfv\'en torsional nodeless oscillations of crustal magneto-active plasma under the action of solely Lorentz restoring force.}
\end{figure}

   The results presented in Fig.9 and Fig.10 show that at indicated input parameters, i.e.,
   the Alfv\'en frequency $\nu_A$ and the fractional depth of seismogenic layer $h=\Delta R/R$,
   the model too adequately reproduces general trends in the data with fairly
   reasonable $\ell$-pole specification of overtones pointed out by integer numbers along x-axis.
   It is seen that the low-frequency QPOs in data for SGR 1806-20, are interpreted as dipole and quadrupole overtones: $\nu(_0a^t_1)=18$ and $\ell(_0a^t_2)=26$ Hz. And the high-frequency kilohertz vibrations with $627$ Hz and $1870$ Hz are unambiguously specified as high-multipole overtones: $\nu(_0a^t_{\ell=42})=627$ Hz and $\nu(_0a^t_{\ell=122})=1870$ Hz.

   \begin{figure}[h]
 \centering{\includegraphics[width=9.0cm]{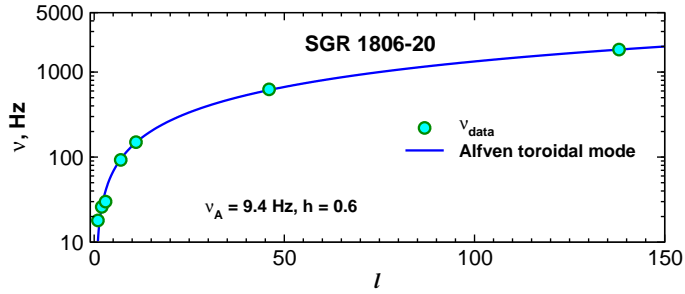}}
 \caption{Same as Fig.9 but for SGRs 1806-20.}
\end{figure}

   However, in this latter scenario of Lorentz-force-dominated vibrations the best fit of data is attained at fairly large value of fractional depth, $h=0.6$, which is much larger than the expected depth of the crust.
   In our opinion, this result may be regarded as indication to that the detected QPOs are formed by coherent vibrations of crustal solid-state plasma and plasma of magnetar corona.

\section{Concluding remarks}

 The neutron stars, both pulsars and magnetars, are seismically active compact objects which are formed, as is widely believed, in the magnetic-flux-conserving supernova collapse of the cores of massive stars coming into existence
 as sources of super strong magnetic field with a fairly complex internal constitution. The internal micro-composition of these compact objects has been and still remains among the most active investigations on search for adequate equations of state of neutron star matter
 (e.g. Weber 1999, Lattimer \& Prakash 2007, Sedrakian 2007, Haensel, Potekhin \& Yakovlev 2007).

 Together with this development, the past three decades have seen increasing recognition that important insight into interior of neutron stars and properties of super dense and ultra magnetized matter provides asteroseimology which seeks to explain fast variability of their luminosity as being produced by seismic vibrations.
 Adhering to this attitude, we have investigated seismic response of core-crust model of quaking paramagnetic neutron star, pictured in Fig.1, by node-free torsional crust-against-core vibrations.
 In this model the source of ultra powerful and highly stable to spontaneous decay magnetic field is
 associated with  massive core, which is thought of as an ultra powerful spherical magnet. The physical
 nature of seismic stability of core-crust coupling is attributed to magnetic cohesion between core and metallic like crust\footnote{At this point it may be appropriate to note that
 considered model permits co-existence of two sources and two compound parts of neutron star magnetism (current-carrying flows in the crust and field-induced spin paramagnetism of core) and consistent
 with the idea about strengthening of magnetic field intensity in young neutron stars, like magnetars, due to thermoelectric effect in the crust (Urpin \& Yakovlev 1980, Blandford et al 1983). And also with
 the idea that damping of crustal currents, thought of as second source of magnetic field of pulsars and magnetars, serves as one of the main factors responsible for the decay of magnetic field of neutron stars (e.g. Goldreich  \& Reisenegger 1992, Pons \& Geppert 2008). It is expected that magnetic field decay should manifest itself in the lengthening of periods of Alfv\'en oscillations.}. The fact that impulsive release of magnetic energy in the starquake, associated with disruption of magnetic field lines on the core-crust interface followed by fracturing the crust by relived magnetic stresses, suggests that model
  under consideration is in line with current understanding of X-ray bursting activity of soft gamma-ray repeaters.
  With this in mind, the obtained spectral formulae have been applied to asteroseismic analysis of data
  on quasi-periodic oscillations of the outburst X-ray emission
   from SGR 1900+14 and SGR 1806-20. In so doing we have investigated two scenarios of post-quake relaxation of paramagnetic neutron star by torsional vibrations of crust against immobile massive core distinguishing restoring forces of vibrations. In first case, the analysis of data on QPOs frequencies has been based on
  assumption that detected QPOs owe their existence to torsional nodeless vibrations restored by joint action of Lorentz magnetic and Hooke's elastic forces and we have found that obtained three-parametric spectral formula provides much better fit of data than two-parametric frequency spectrum of the early considered model of global magneto-elastic vibrations (Bastrukov et al 2009a).
  In second case of vibrations restored by solely Lorentz force of magnetic field stresses we found that obtained two-parametric frequency spectrum can too be fairly reasonably reconciled with detected QPOs frequencies.
  With all above in mind we conclude that main part in quake-induced torsional vibrations of crustal solid-state plasma about
  axis of magnetic field frozen in the immobile core plays Lorentz restoring force of magnetic field stresses and that observed frequencies of QPOs can be interpreted as an indirect evidence for permanent magnetization of super dense matter in the deep interior of magnetars.

 The authors are grateful to Dima Podgainy (JINR, Dubna) for helpful assistance, Kinwah Wu for discission
 and Ziri Younsi (Mullard Space Science Laboratory, University College London) for critical reading of the manuscript on the initial stage of this work. This work is a part of projects on investigation of variability of high-energy emission from compact sources supported by NSC of Taiwan, grant numbers  NSC-96-2628-M-007-012-MY3 and NSC-97-2811-M-007-003.

\end{document}